\documentclass[twocolumn,10pt,amsmath,amssymb,prl]{revtex4-1}
\usepackage{textcomp}
\usepackage{graphicx}
\usepackage{longtable}
\usepackage{color}
\usepackage{bm}
\usepackage{dcolumn}
\usepackage{epstopdf}
\usepackage{hyperref}

\newcommand{\cmi}[1]
{
  cm$^{-1}$#1
}
\newcommand{\EE}[0]
{
  \mathcal{E}
}
\newcommand{\abin}[0]
{
  \textit{ab initio}
}

\begin{document}

\title{Ultracold rare-earth magnetic atoms with an electric dipole moment}

\author{Maxence Lepers$^{1,2}$, Hui Li$^{1}$, Jean-Fran{\c c}ois Wyart$^{1,3}$, Goulven Qu{\'e}m{\'e}ner$^{1}$ and Olivier Dulieu$^{1}$}
\address{${}^{1}$Laboratoire Aim{\'e} Cotton, CNRS, Universit{\'e} Paris-Sud, ENS Paris-Saclay, Universit{\'e} Paris-Saclay, 91405 Orsay, France}
\address{${}^{2}$Laboratoire Interdisciplinaire Carnot de Bourgogne, CNRS, Universit\'e de Bourgogne Franche-Comt\'e, 21078 Dijon, France}
\email{maxence.lepers@u-bourgogne.fr}
\address{${}^{3}$LERMA, Observatoire de Paris-Meudon, PSL Research University, Sorbonne Universit{\'e}s, UPMC Univ.~Paris 6, CNRS UMR8112, 92195 Meudon, France}

\date{\today}

\begin{abstract}
We propose a new method to produce an electric and magnetic dipolar gas of ultracold dysprosium atoms. The pair of nearly degenerate energy levels of opposite parity, at 17513.33 cm$^{-1}$ with electronic angular momentum $J=10$, and at 17514.50 cm$^{-1}$ with $J=9$, can be mixed with an external electric field, thus inducing an electric dipole moment in the laboratory frame. For field amplitudes relevant to current-day experiments, we predict a magnetic dipole moment up to 13 Bohr magnetons, and an electric dipole moment up to 0.22 Debye, which is similar to the values obtained for alkali-metal diatomics. When a magnetic field is present, we show that the electric dipole moment is strongly dependent on the angle between the fields. The lifetime of the field-mixed levels is found in the millisecond range, thus allowing for suitable experimental detection and manipulation.
\end{abstract}

\maketitle

\paragraph{Introduction.}

In a classical neutral charge distribution, a dipole moment appears with a separation between the barycenter of positive and negative charges \cite{jackson1999}. An obvious example is provided by an heteronuclear diatomic molecule, which possesses a permanent dipole moment along its interatomic axis. It will manifest in the laboratory frame when such a molecule is placed in an external electric field, acquiring a preferred orientation along the direction of the field. Moreover, a neutral atom placed in an external electric field acquires a small dipole moment, as the spherical symmetry of space is broken. This effect is spectacularly maximized in Rydberg atoms, where the induced dipole moment scales as $n^2$, where $n$ is the principal quantum number of the considered Rydberg state \cite{gallagher2005}.

At the single-particle scale, the external electric field mixes even and odd-parity levels of the energy spectrum: rotational levels for a diatomic molecule (see \textit{e.g.}~\cite{brieger1984}), or levels with different orbital angular momenta for Rydberg atoms (see \textit{e.g.}~\cite{zimmerman1979}). In both cases, this leads to a pronounced linear Stark shift on the energy levels, revealing the existence of a permanent dipole moment in the laboratory frame. More surprisingly, it has been observed that a homonuclear diatomic molecule can exhibit a permanent dipole moment in the laboratory frame, when it combines a ground state atom bound inside the spatial extension of a Rydberg atom \cite{greene2000, li2011}.

The search for such dipolar systems, involving especially lanthanide atoms, is currently very active in the context of ultracold dilute gases \cite{lu2010, sukachev2010, aikawa2012, miao2014, frisch2015, kadau2016, dreon2016, lucioni2017, ulitzsch2017, becher2018, ravensbergen2018}. Indeed, the particles of the gas interact through a highly anisotropic long-range potential energy varying as the inverse cubic power of their spatial separation \cite{stone1996, lepers2017}. Prospects related to many-body physics, quantum simulation and ultracold chemistry are nowadays within reach experimentally \cite{baranov2008, lahaye2009, baranov2012}. A particular attention is paid on gases with an electric and a magnetic dipole moment, which up to now consist of paramagnetic polar diatomics \cite{zuchowski2010a, pasquiou2013, barry2014, khramov2014, tomza2014, zuchowski2014, karra2016, quemener2016, reens2017, rvachov2017}.

In this Letter, we propose a new method to produce an electric and magnetic dipolar gas of ultracold dysprosium atoms. Our method is based on the electric-field mixing of quasi-degenerate opposite-parity energy levels, which appear accidentally in the rich spectra of lanthanides. Historically, the pair of levels at 19797.96 cm$^{-1}$ with electronic angular momenta $J=10$ has been employed for fundamental measurements  \cite{budker1993, cingoz2007, leefer2013}. However, their reduced transition dipole moment, equal to 0.015 atomic units (a.u.) \cite{budker1994}, is not sufficient to observe dipolar effects. On the contrary, the odd-parity level $|a\rangle$ at $E_a=17513.33$ cm$^{-1}$ with $J_a=10$ and the even-parity level $|b\rangle$ at $E_b=17514.50$ cm$^{-1}$ with $J_b=9$, which present a reduced transition dipole moment of $3.21$ a.u., are very promising for dipolar gases \cite{wyart1974}.

We calculate the energies, electric (EDMs) and magnetic dipole moments (MDMs) of a dysprosium atom in a superposition of levels $|a\rangle$ and $|b\rangle$, and submitted to an electric and a magnetic field with an arbitrary respective orientation. For field amplitudes relevant to current-day experiments, we predict a MDM of $\mu_\mathrm{max}=13$ Bohr magnetons, to our knowledge the largest value observed in ultracold experiments, and an EDM of $d_\mathrm{max}=0.22$ Debye, which is similar to the values of diatomic molecules \cite{ni2010}. We also demonstrate a strong control of the electric dipole moment, which ranges from 0 to $d_\mathrm{max}$ as a function of the angle between the fields. Because $|a\rangle$ and $|b\rangle$ are excited levels, we also calculate the atomic radiative lifetime as functions of the fields parameters, and obtain a few millisecond for the level characterized by $\mu_\mathrm{max}$ and $d_\mathrm{max}$. Finally, we show that our method is applicable for all bosonic and fermionic isotopes.

\paragraph{Model.}

We consider an atom lying in two energy levels $|a\rangle$ and $|b\rangle$, of energies $E_i$ and total angular momentum $J_i$ ($i=a,\,b$). Firstly, we consider bosonic isotopes which have no nuclear spin, $I=0$. In absence of field, each level $|i\rangle$ is $(2J_i+1)$-time degenerate, and the corresponding Zeeman subslevels are labeled with their magnetic quantum number $M_i$. The atom is submitted both to a magnetic field $\mathbf{B} = B\mathbf{e}_z$, with $\mathbf{e}_z$ the unit vector in the $z$ direction, taken as quantization axis, and to electric field $\mathbf{E} = \EE\mathbf{u}$, with $\mathbf{u}$ a unit vector in the direction given by the polar angles $\theta$ and $\phi=0$. In the basis $\{|M_a = -J_a\rangle, \,..., \, |+J_a\rangle$, $|M_b = -J_b\rangle, \,...,\, |+J_b\rangle\}$ spanned by the Zeeman sublevels of $|a\rangle$ and $|b\rangle$, the Hamiltonian can be written
\begin{equation}
  \hat{H} = \sum_{i=a,b} E_i \sum_{M_i=-J_i}^{J_i} 
    |M_i\rangle\langle M_i| + \hat{W}_Z + \hat{W}_S \,.
  \label{eq:hamilt}
\end{equation}
The Zeeman Hamiltonian $\hat{W}_Z$ only contains diagonal terms equal to $M_i g_i \mu_B B$, with $g_i$ the Land\'e g-factor of level $|i\rangle$. The last term of Eq.~\eqref{eq:hamilt} is the Stark Hamiltonian, which couples sublevels $|M_a\rangle$ with sublevels $|M_b\rangle$ as
\begin{align}
  \langle M_a| \hat{W}_S |M_b\rangle & = -\sqrt{\frac{4\pi}{3(2J_a+1)}}
    \langle a\Vert\hat{\mathbf{d}}\Vert b\rangle \, \EE
  \nonumber \\
    & \times Y_{1,M_a-M_b}^*(\theta,0) \, C_{J_bM_b,1,M_a-M_b}^{J_aM_a} \,,
  \label{eq:ws}
\end{align}
where $\langle a\Vert\hat{\mathbf{d}}\Vert b\rangle$ is the reduced transition dipole moment, $Y_{kq}(\theta,\phi)$ a spherical harmonics and $C_{a\alpha b\beta}^{c\gamma}$ a Clebsch-Gordan coefficient \cite{varshalovich1988}. For given values of $\EE$, $B$ and $\theta$, we calculate the eigenvalues $E_n$ and eigenvectors
\begin{equation}
  |\Psi_n\rangle = \sum_{i=a,b} \sum_{M_i=-J_i}^{J_i}
    c_{n,M_i} |M_i\rangle
\end{equation}
of the Hamiltonian in Eq.~\eqref{eq:hamilt}.

The energy levels that we consider here are $E_a=17513.33$ \cmi{,} $J_a=10$ and $E_b=17514.50$ \cmi{,} $J_b=9$. Their Land\'e g-factors $g_a=1.30$ and $g_b=1.32$ are experimental values taken from Ref.~\cite{NIST_ASD}. The reduced transition dipole moment $\langle a\Vert\hat{\mathbf{d}}\Vert b\rangle$ is calculated using the method developed in our previous works \cite{lepers2014, lepers2016, li2017, li2017a}. Firstly, odd-level energies are taken from Ref.~\cite{li2017} which include the electronic configurations $\mathrm{[Xe]}4f^{10}6s6p$ and $\mathrm{[Xe]}4f^95d6s^2$, [Xe] being the xenon core. Even-level energies are calculated with the configurations $\mathrm{[Xe]}4f^{10}6s^2$, $\mathrm{[Xe]}4f^{10}5d6s$ and $\mathrm{[Xe]}4f^96s^26p$ \cite{wyart2018}. Secondly, following Ref.~\cite{li2017}, we adjust the mono-electronic transition dipole moments by multiplying their\abin values by appropriate scaling factors \cite{lepers2016}, equal to 0.794 for $\langle 6s|\hat{r}|6p\rangle$, 0.97 for $\langle 4f|\hat{r}|5d\rangle$ and 0.80 for $\langle 5d|\hat{r}|6p\rangle$. From the resulting Einstein coefficients, we can extract $\langle a\Vert\hat{\mathbf{d}}\Vert b\rangle = 3.21$ a.u., as well as the natural linewidth $\gamma_b = 2.98 \times 10^{4}$ s$^{-1}$. At the electric-dipole approximation, $\gamma_a$ vanishes; considering electric-quadrupole and magnetic-dipole transitions, it can be estimated with the Cowan codes \cite{cowan1981} as $\gamma_a = 3.56 \times 10^{-2}$ $s^{-1}$.

\paragraph{Energies in electric and magnetic fields.}


\begin{figure}
  \includegraphics[width=8cm]{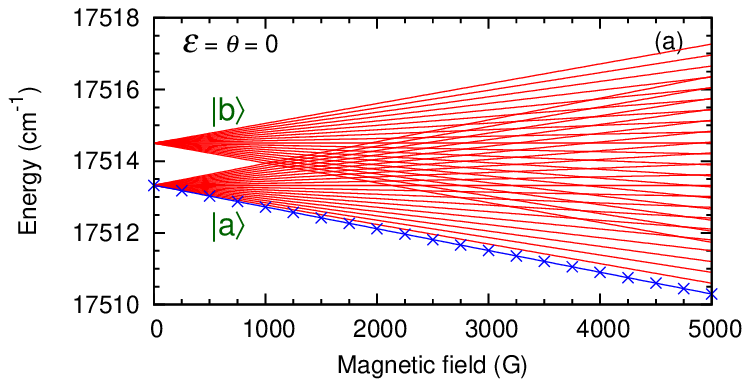}
  \includegraphics[width=8cm]{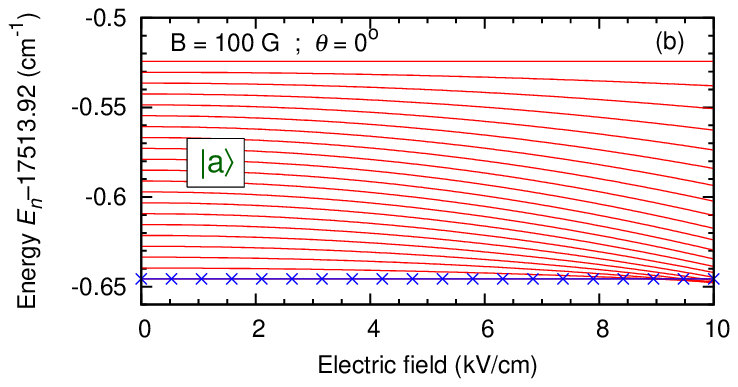}
  \includegraphics[width=8cm]{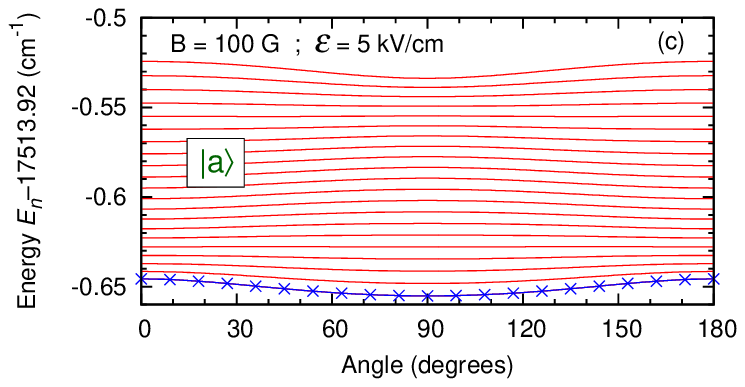}
  \caption{\label{fig:ener} Eigenvalues of the atom-field Hamiltonian \eqref{eq:hamilt} as functions of: (a) the magnetic field $B$ for vanishing electric field and angle $\EE = \theta = 0$; (b) the electric field $\EE$ for $B = 100$ G and $\theta = 0^\circ$; (c) the angle $\theta$ for $B = 100$ G and $\EE = 5$ kV/cm. In panels (b) and (c), the origin of energies is taken at $(E_a+E_b)/2 = 17513.92$ \cmi{.} The blue curve with crosses corresponds to the eigenstate converging towards $|M_a = -10 \rangle$ when $\theta\to 0$.}
\end{figure}

Figure \ref{fig:ener}(a) shows the eigenvalues of the Hamiltonian \eqref{eq:hamilt} as functions of the magnetic field for $\EE = \theta = 0$. The field splits levels $|a\rangle$ and $|b\rangle$ into 21 and 19 sublevels respectively, each one associated with a given $M_a$ or $M_b$. On fig.~\ref{fig:ener}(a), we emphasize the lowest sublevel $|M_a=-10\rangle$, in which ultracold atoms are usually prepared. Due to the close Land\'e g-factors, the two Zeeman manifolds look very similar, \textit{i.e.}~the branches characterized by the same values $M_a=M_b$ are almost parallel. For $B \ge 1000$ Gauss, the two Zeeman manifolds overlap; but because the magnetic field conserves parity, the sublevels of $|a\rangle$ and $|b\rangle$ are not mixed. Provoking that mixing is the role of the electric field. 

On figure \ref{fig:ener}(b), we plot the 21 lowest eigenvalues of Eq.~\eqref{eq:hamilt} as functions of the electric field for $B = 100$ Gauss and $\theta = 0^\circ$. We focus on the eigenstates converging to the sublevels of $|a\rangle$ when $\EE\to 0$. In the range of field amplitudes chosen in Figs.~\ref{fig:ener}(a) and (b), which corresponds to current experimental possibilities, the influence of $\EE$ is much weaker than the influence of $B$. On Fig.~\ref{fig:ener}(b), the energies decrease quadratically with the electric field, because the sublevels of $|a\rangle$ are repelled by the sublevels of $|b\rangle$. Since $\theta = 0^\circ$, the $z$ component of the total angular momentum is conserved, and so, the sublevels for which $M_a=M_b$ are coupled in pairs. In consequence, the sublevels $|M_a = \pm 10 \rangle$ are insensitive to the electric field, as they have no counterparts among the sublevels of $|b\rangle$ (recalling that $J_b=9$). 

The only way to couple the $|M_a = \pm 10 \rangle$ sublevels to the other ones is to rotate, say, the electric field, and thus break the cylindrical symmetry around the $z$ axis. On figure \ref{fig:ener}(c), the 21 lowest eigenvalues of Eq.~\eqref{eq:hamilt} are now shown as function of the angle $\theta$, for fixed field amplitudes, $\EE = 5$ kV/cm and $B = 100$ Gauss. Even if the corresponding eigenvectors are not associated with a single sublevel $|M_i\rangle$ (unlike Figs.~\ref{fig:ener}(a) and (b)), they can conveniently be labeled $|\overline{M}_i\rangle$ after their field-free or $\theta=0$ counterparts. For a given eigenstate, the $\theta$-dependence of energy is weak. However for $|\overline{M}_a =\pm 10\rangle$, the energy decrease reveals the repulsion with sublevels of $|b\rangle$, which is maximum for $\theta = 90^\circ$.

\paragraph{Magnetic and electric dipole moments.}

The $z$ component of the MDM associated with the eigenvector $|\Psi_n\rangle$ is equal to 
\begin{equation}
  \mu_n = -\mu_B \sum_{i=a,b} g_i
    \sum_{M_i=-J_i}^{J_i} \left|c_{n,M_i}\right|^2 M_i .
  \label{eq:dip-mag}
\end{equation}
Since the eigenvectors are mostly determined by their field-free counterparts, $\mu_n$ does not change significantly in our range of field amplitudes; it is approximately equal to $\mu_n \approx -\overline{M}_a g_a\mu_B$ for $n\in[1;21]$ and $\mu_n \approx -\overline{M}_b g_b\mu_B$ for $n\in[22;40]$. For instance, the state $|\overline{M}_a = -10\rangle$ has the maximal value $\mu_\mathrm{max} = 13.0 \times \mu_B$.


\begin{figure}
  \includegraphics[width=8cm]{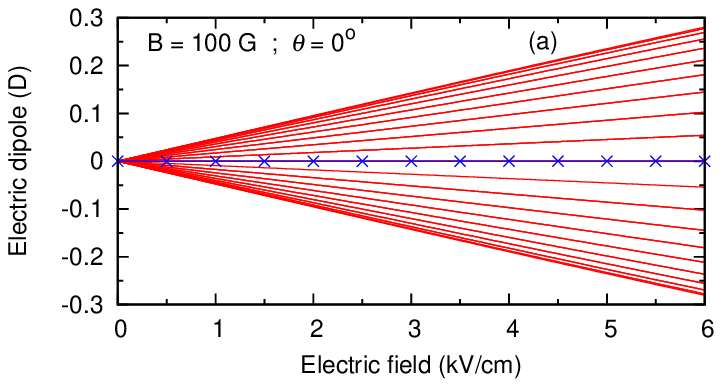}
  \includegraphics[width=8cm]{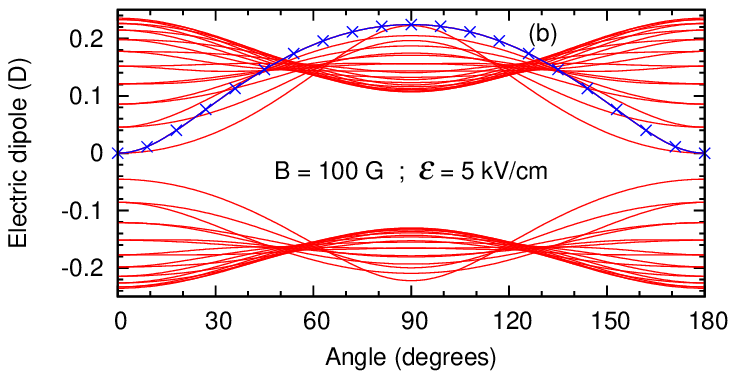}
  \caption{\label{fig:dip} Electric dipole moments, see Eq.~\eqref{eq:dip-elec}, associated with the eigenstates of the atom-field Hamiltonian \eqref{eq:hamilt} as functions of: (a) the electric field $\EE$ for $B = 100$ G and $\theta = 0^\circ$; (b) the angle $\theta$ for $B = 100$ G and $\EE = 5$ kV/cm.  The blue curve with crosses corresponds to the eigenstate $|\overline{M}_a = -10 \rangle$.}
\end{figure}

The mean EDM $d_n = \langle\Psi_n| \mathbf{d}\cdot\mathbf{u} |\Psi_n\rangle$ associated with the eigenvector $|\Psi_n\rangle$ in the direction $\mathbf{u}$ of the electric field is
\begin{equation}
  d_n = -\frac{1}{\EE} \sum_{M_a,M_b} c_{n,M_a}^* c_{n,M_b}
    \left\langle M_a\right| \hat{W}_S \left|M_b \right\rangle + c.c. \,,
  \label{eq:dip-elec}
\end{equation}
where the matrix element of $\hat{W}_S$ is given in Eq.~\eqref{eq:ws}. Figure \ref{fig:dip}(a) presents the EDMs as functions of the electric field $\EE$, for $B = 100$ G and $\theta = 0 ^\circ$. In this case, the graph is symmetric about the $y$ axis. All the curves vary linearly with $\EE$; all, except the lowest and highest ones, correspond to two eigenstates. In agreement with Fig.~\ref{fig:ener}(b), the curve $d_n=0$ is associated with $|\overline{M}_a = \pm 10 \rangle$ ($n=1$ and 21). The lowest and highest curves belong to $\overline{M}_{a,b} = 0$, for which by contrast, the MDM vanishes.

As shows figure \ref{fig:dip}(b), the EDMs change dramatically as function of the angle $\theta$.  In particular, the EDM of the eigenstate $|\overline{M}_a=-10 \rangle$ ranges continuously from 0 to a maximum $d_\mathrm{max} = 0.224$ Debye for $\theta = 90 ^\circ$. The eigenstate $|\overline{M}_a = 10 \rangle$ follows a similar evolution, except that its curve is sharper around its maximum. In contrast, the EDM of the eigenstate $|\overline{M}_a=0\rangle$, which is the largest for $\theta = 0 ^\circ$, becomes the smallest for $90^\circ$. Compared to the eigenstates $|\overline{M}_a\rangle$, the curves corresponding to the eigenstates $|\overline{M}_b\rangle$ exhibit an approximate reflection symmetry around the $y$ axis. Finally, it is important to mention that the influence of the magnetic field on the EDMs is weak in the amplitude range of Fig.~\ref{fig:ener}(a).

\paragraph{Radiative lifetimes.}


\begin{figure}
  \includegraphics[width=8cm]{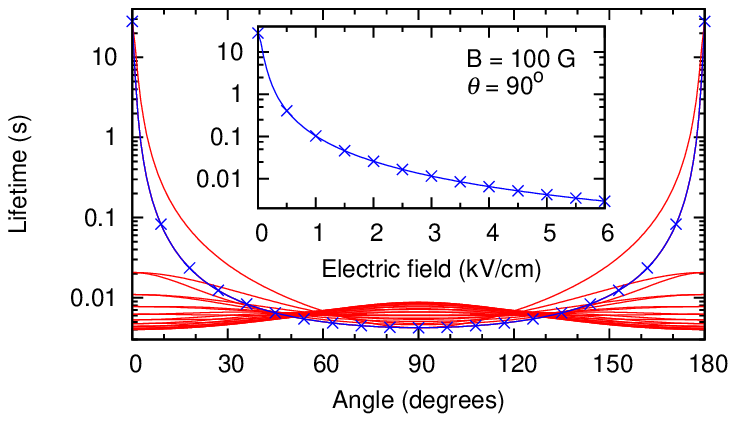}
  \caption{\label{fig:lftm} Radiative lifetimes, see Eq.~\eqref{eq:lifetm}, associated with the eigenstates $|\overline{M}_a\rangle$ as functions of the angle $\theta$ for $B = 100$ G and $\EE = 5$ kV/cm. The blue curve with crosses corresponds to the eigenstate $|\overline{M}_a = -10 \rangle$. The inset shows the lifetime of this eigenstate as a function of $\EE$ for $B = 100$ G and $\theta = 90 ^\circ$.}
\end{figure}

The radiative lifetime $\tau_n = 1/\gamma_n$ associated with eigenvector $|\Psi_n\rangle$ is such that $\gamma_n$ is an arithmetic average of the natural line widths of $|a\rangle$ and $|b\rangle$,
\begin{equation}
  \tau_n = \gamma_n^{-1} = \left( \sum_{i=a,b} \gamma_i 
    \sum_{M_i=-J_i}^{J_i} \left|c_{n,M_i}\right|^2 \right)^{-1} .
  \label{eq:lifetm}
\end{equation}
Figure \ref{fig:lftm} displays the lifetimes of all eigenstates of Eq.~\eqref{eq:hamilt} as functions of the angle $\theta$ for $\EE=5$ kV/cm and $B=100$ G. Because the natural line widths $\gamma_a$ and $\gamma_b$ differ by 6 orders of magnitude, the lifetimes $\tau_n$ are also spread over a similar range. At the field amplitudes of Fig.~\ref{fig:lftm}, the eigenvectors $|\overline{M}_a\rangle$ are composed at least of $90$ \% of sublevels of $|a\rangle$, and similarly for eigenstates $|\overline{M}_b \rangle$. Therefore, the lifetimes of eigenstates $|\overline{M}_b \rangle$ (not shown on Fig.~\ref{fig:lftm}) are approximately $1/\gamma_b$, and they weakly depend on $\theta$.
As for the eigenstates $|\overline{M}_a \rangle$, their small $|b\rangle$ components, say $\varepsilon$, induces lifetimes roughly equal to $\approx \tau_b/\varepsilon^2$. For $|\overline{M}_a =\pm 10 \rangle$, the lifetime ranges from $\tau_\mathrm{min} = 4.22$ ms for $\theta = 90 ^\circ$ to $\tau_a = 28.1$ s for $\theta = 0 ^\circ$. Again, this illustrates that the coupling with the sublevels of $|b\rangle$ is maximum for perpendicular fields and absent for colinear ones.

The inset of figure \ref{fig:lftm} shows the lifetime of the eigenstate $|\overline{M}_a=-10\rangle$ as function of $\EE$. In this range of field amplitude, $\tau_1$ scales as $\EE^{-2}$. So, a large amplitude $\EE$ can strongly affect the lifetime of the atoms; but on the other hand, $\EE$ needs to be sufficient to induce a noticeable EDM. So, there is a compromise to find between EDM and lifetime, by tuning the electric-field amplitude and the angle between the fields.

\paragraph{Fermionic isotopes.}

There are two fermionic isotopes of dysprosium, $^{161}$Dy and $^{163}$Dy, both with a nuclear spin $I=5/2$. A given hyperfine sublevel is characterized by the total (electronic+nuclear) angular momentum $F_i$ and its $z$-projection $M_{F_i}$, where $|J_i-I| \le F_i\le J_i+I$, and $-F_i \le M_{F_i} \le F_i$. Namely, $F_a$ ranges from $15/2$ to $25/2$, and $F_b$ ranges from $13/2$ to $23/2$. The hyperfine sublevels are constructed by angular-momentum addition of $\mathbf{J}_i$ and $\mathbf{I}$, \textit{i.e.}~$ |F_iM_{F_i}\rangle = \sum_{M_iM_I} C_{J_iM_iIM_I}^{F_iM_{F_i}} |J_iM_i\rangle |IM_I\rangle$. Compared to Eq.~\eqref{eq:hamilt}, the Hamiltonian $\hat{H}'$ is modified as
\begin{equation}
  \hat{H}' = \sum_{i=a,b} \sum_{F_i}
    E_{F_i} \sum_{M_{F_i}} |F_iM_{F_i} \rangle
    \langle F_iM_{F_i}| + \hat{W}_Z + \hat{W}_S
  \label{eq:hamilt-ferm}
\end{equation}
where $E_{F_i}$ is the hyperfine energy depending on the magnetic-dipole and electric-quadrupole constants $A_i$ and $B_i$. For $^{163}$Dy, they have been calculated in Ref.~\cite{dzuba1994}: $A_a = 225$ MHz, $B_a = 2434$ MHz, $A_b = 237$ MHz and $B_b = 706$ MHz. For $^{161}$Dy, we apply the relations $(^{161}A_i) = -0.714\times (^{163}A_i)$ and $(^{161}B_i) = 0.947\times (^{163}B_i)$ given in Ref.~\cite{eliel1980}. The matrix elements of the Zeeman $\hat{W}_Z$ and Stark Hamiltonians $\hat{W}_S$ are calculated by assuming that they do not act on the nuclear quantum number $M_I$, and by using the formulas without hyperfine structure (see Eq.~\eqref{eq:ws} and text above).


\begin{table}
  \caption{\label{tab:isot} Maximal electric dipole moment $d_\mathrm{max}$, dipolar length $a_\mathrm{d}$ \cite{julienne2011} and minimal lifetime $\tau_\mathrm{min}$ obtained for different isotopes of dysprosium for an electric field $\EE = 5$ kV/cm, a magnetic field $B = 100$ G and an angle $\theta = 90 ^\circ$. The results of $^{162}$Dy are also valid for the other bosonic isotopes $^{156}$Dy, $^{158}$Dy, $^{160}$Dy and $^{164}$Dy.}
  \begin{ruledtabular}
  \begin{tabular}{lcccc}
     & $d_\mathrm{max}$ (D) & $a_\mathrm{d}$ (a.u.)
     & $\tau_\mathrm{min}$ (ms) \\
    \hline 
     $^{161}$Dy$\phantom{A^{A^A}}$ & 0.225 & 2299 & 4.18 \\
     $^{162}$Dy & 0.224 & 2293 & 4.22 \\
     $^{163}$Dy & 0.222 & 2266 & 4.23 \\ 
  \end{tabular} 
  \end{ruledtabular}
\end{table}

After diagonalizing Eq.~\eqref{eq:hamilt-ferm}, one obtains 240 eigenstates (compared to 40 in the bosonic case). Despite their large number of curves, the plots of energies, EDMs and lifetimes show similar features to figures \ref{fig:ener}--\ref{fig:lftm}. The eignestates $|\Psi'_n\rangle$ can be labeled $|\overline{F}_i \overline{M}_{F_i} \rangle$ after their field-free counterparts $|F_iM_{F_i}\rangle$. Moreover, the {}``stretched'' eigenstates $|\overline{F}_a \overline{M}_{F_a}\rangle = |25/2,\pm 25/2\rangle$ are not sensitive to the electric field for $\theta = 0^\circ$, and maximally coupled for $\theta = 90 ^\circ$; and so, their EDMs range from 0 up to $d_\mathrm{max}$ and their lifetimes from $\tau_a$ down to $\tau_\mathrm{min}$. As shows Table \ref{tab:isot}, for the same field characteristics, the values of $d_\mathrm{max}$ and $\tau_\mathrm{min}$ are very similar from one isotope to another.

Table \ref{tab:isot} also contains the so-called dipolar length $a_\mathrm{d} = md_\mathrm{max}^2/\hbar^2$ \cite{julienne2011}. It characterizes the length at and beyond which the dipole-dipole interaction between two particles is dominant. For the $^{161}$Dy isotope, one can reach a dipolar length of $a_\mathrm{d} = 2299$ a.u.. To compare with, at $\EE = 5$ kV/cm and an induced dipole moment of $~0.22$ D \cite{ni2010}, $^{40}$K$^{87}$Rb has a length of $a_\mathrm{d} = 1734$ a.u.. Similarly, a length of $a_\mathrm{d} = 1150$ a.u. was reached \cite{frisch2015} for magnetic dipolar Feshbach molecules of $^{168}$Er$_2$. With the particular set-up of electric and magnetic fields employed in this study, we show that one can reach comparable and even stronger dipolar character with atoms in excited states than with certain diatomic molecules.

\paragraph{Conclusion.}

We have demonstrated the possibility to induce a strong electric dipole moment on atomic dysprosium, in addition to its large magnetic dipole moment. To do so, the atoms should be prepared in a superposition of nearly degenerate excited levels using an electric and a magnetic field of arbitrary orientations. We show a remarkable control of the electric dipole moment and radiative lifetime by tuning the angle between the fields.
Since the two levels are metastable, they are not accessible by one-photon transition from the ground level. Instead, one could perform a Raman transition between the ground level $|g\rangle$ ($J_g=8$) and the level $|b \rangle$ ($J_b=9$) of leading configuration $\mathrm{[Xe]}4f^{10}5d6s$, through the upper levels at 23736.61, 23832.06 or 23877.74 \cmi{,} whose $\mathrm{[Xe]}4f^{10}6s6p$ character insures significant transition strengths with $|g\rangle$ and $|b\rangle$. In the spectrum of other lanthanides, there exist pairs of quasi-degenerate levels accessible from the ground state, for instance the levels at 24357.90 and 24660.80 \cmi{} in holmium, but in turn their radiative lifetime is much shorter \cite{den-hartog1999}.

\paragraph{Acknowledgments.}

We acknowledge support from {}``DIM Nano-K'' under the project {}``InterDy'', and from {}``Agence Nationale de la Recherche'' (ANR) under the project {}``COPOMOL'' (contract ANR-13-IS04-0004-01). We also acknowledge the use of the computing center {}``M{\'e}soLUM'' of the LUMAT research federation (FR LUMAT 2764).


%

\end{document}